# COSMIC-RAY ORIGIN IN OB ASSOCIATIONS AND PREFERENTIAL ACCELERATION OF REFRACTORY ELEMENTS: EVIDENCE FROM ABUNDANCES OF ELEMENTS $_{26}$Fe THROUGH $_{34}$Se.

B. F. Rauch[1], J. T. Link[1,2], K. Lodders[1], M. H. Israel[1], L. M. Barbier[3], W. R. Binns[1], E. R. Christian[3], J. R. Cummings[1,3], G. A. de Nolfo[2], S. Geier[4,7], R. A. Mewaldt[4], J. W. Mitchell[3], S. M. Schindler[4], L. M. Scott[1,8], E. C. Stone[4], R. E. Streitmatter[3], C. J. Waddington[5], and M. E. Wiedenbeck[6]

[1]Washington University, St. Louis, MO 63130, USA
[2]GSFC/CRESST, Greenbelt, MD 20771, USA
[3]NASA Goddard Space Flight Center, Greenbelt, MD 20771, USA
[4]California Institute of Technology, Pasadena, CA 91125, USA
[5]University of Minnesota, Minneapolis, MN 55455, USA
[6]Jet Propulsion Laboratory, California Institute of Technology, Pasadena, CA 91109, USA
[7]Current address: Jet Propulsion Laboratory, Calif. Inst. of Tech., Pasadena, CA 91109, USA
[8]Current address: Rutgers University, Piscataway, NJ 08854, USA

Corresponding author:

Martin H. Israel
Campus Box 1105
Washington University
One Brookings Drive
St. Louis, MO 63130

e-mail: mhi@wustl.edu

## ABSTRACT

We report abundances of elements from $_{26}$Fe to $_{34}$Se in the cosmic radiation measured during fifty days of exposure of the Trans-Iron Galactic Element Recorder (TIGER) balloon-borne instrument. These observations add support to the concept that the bulk of cosmic-ray acceleration takes place in OB associations, and they further support cosmic-ray acceleration models in which elements present in interstellar grains are accelerated preferentially compared with those found in interstellar gas.

## 1. INTRODUCTION

The observed elemental and isotopic abundances of cosmic-ray nuclei provide evidence concerning the regions where the bulk of the Galactic cosmic rays (GCR) is accelerated. After correcting for the effects of fragmentation due to nuclear interactions in the interstellar medium, one can derive the composition of the accelerated cosmic rays and search for signatures indicative of particular Galactic environments.

Measurements of the abundance ratio $^{22}$Ne/$^{20}$Ne (Garcia-Munoz et al. 1979, Wiedenbeck & Greiner 1981, Mewaldt et al. 1980, Lukasiak et al. 1994, Webber et al.

1997, Connell & Simpson 1997, DuVernois et al. 1996, Binns et al. 2005) showed that this ratio in the cosmic-ray source is substantially higher than in the solar system. The most widely accepted explanation for the high relative abundance of $^{22}$Ne in the cosmic rays is the presence in the cosmic-ray source of a substantial fraction of ejecta from Wolf-Rayet stars (Cassé & Paul 1982, Prantzos et al. 1987) mixed with material that has abundances similar to those in the solar system.

Higdon & Lingenfelter (2003) noted that an elevated $^{22}$Ne/$^{20}$Ne ratio (from Binns et al. 2001) is the natural consequence of the fact that most supernovae, the accepted source of cosmic-ray energy, and most Wolf-Rayet stars occur in OB associations. They fit the measured cosmic-ray $^{22}$Ne/$^{20}$Ne ratio with a cosmic-ray source mixture in superbubbles composed of $(18 \pm 5)\%$ by mass of ejecta from Wolf-Rayet stars and core-collapse supernovae and the remaining material of composition similar to that in the Solar-System.

Binns et al. (2005) used data from the Advanced Composition Explorer/Cosmic Ray Isotope Spectrometer (ACE/CRIS) instrument (Stone et al. 1998) to obtain a precise determination of the $^{22}$Ne/$^{20}$Ne ratio. When corrected back to the cosmic-ray source this ratio was found to be a factor of $5.3 \pm 0.3$ above the solar-system value. They also noted that among twelve additional isotopic abundance ratios, calculated at the GCR source, reported from ACE/CRIS (see, e.g. Wiedenbeck et al. 2007 and references therein) only the $^{58}$Fe/$^{56}$Fe ratio differed significantly from the solar ratio, with an enhancement factor $1.69 \pm 0.27$. They demonstrated that all these isotope ratios, as well as the element ratios C/O, N/O, and N/Ne are consistent with a mix of approximately 20% Wolf-Rayet ejecta and 80% normal matter with solar-system composition, when adjusted for physical and chemical fractionation properties, discussed in the following paragraph.

In addition to the signature of the site of cosmic-ray acceleration carried by the cosmic-ray isotopic and elemental abundances, the elemental cosmic-ray composition points to atomic properties of the various elements that affect the acceleration process. Cassé & Goret (1978) noted that elements with first ionization potential (FIP) $<\sim 10$ eV were more abundant in the cosmic-ray source, relative to solar-system abundances, than elements with higher FIP. Epstein (1980) and Cesarsky & Bibring (1981) noted that similar elemental abundance variations could be explained by a higher abundance of refractory than volatile elements. More recently Meyer et al. (1997) and Ellison et al. (1997) gave further support to a model in which refractory elements are enriched in the cosmic rays, and Higdon et al. (1998) showed that sputtering from a mix of grains from massive star ejecta and interstellar medium could account for the enhanced abundance of the refractory elements. In the Ellison et al. (1997) model, dust grains become charged by photoionization and accelerated to moderate energies by supernova shocks. Atoms that are sputtered off of these grains have suprathermal energies and are accelerated more efficiently to cosmic-ray energies than are atoms originating in the thermal interstellar gas. *[Note: The published ApJ version of this article implied an improper attribution for the model described in the previous two sentences. The version of this article presented here corrects that error to make clear that the model comes from Ellison et al (1997).]*

In this paper, we present analysis of measurements of the elemental composition of cosmic rays with atomic number $26 \leq Z \leq 38$ made by the Trans-Iron Galactic Element Recorder (TIGER) instrument during two high-altitude balloon flights over Antarctica. The first launch was on 2001 December 21; the instrument floated for 32 days at an average altitude of 36 km (5.5 mb pressure). The second launch was on 2003 December 17; the float period lasted 18 days at an average altitude of 39 km (4.1 mb pressure).

These flights allowed the best measurements to date of the abundances relative to $_{26}$Fe of $_{31}$Ga, $_{32}$Ge, and $_{34}$Se, as well as an upper limit to the abundance of $_{33}$As and a preliminary measurement of the abundance of $_{38}$Sr. We demonstrate that the nearly equal abundances of Ga and Ge cannot be explained by a cosmic-ray source of solar-system composition, but that the abundances of these and other elements are consistent with a cosmic-ray source mixture of about 20% ejecta of massive stars (including Wolf-Rayet stars and core-collapse supernovae) mixed with 80% material of solar-system composition, and an acceleration mechanism in which refractory elements are preferentially accelerated over volatile elements by about a factor of three to four.

## 2. INSTRUMENT DESCRIPTION

The TIGER instrument (Figure 1) is composed of two plastic scintillators (S1 and S2) near the top of the instrument and two (S3 and S4) near the bottom; two Cherenkov detectors, one with an aerogel radiator (C0) and one with an acrylic radiator (C1); and a scintillating fiber hodoscope consisting of an *x,y* plane near the top of the instrument and an *x,y* plane near the bottom.

Each of the scintillators is a 0.8-cm-thick sheet of 116 cm x 116 cm polyvinyl toluene plastic scintillator (St. Gobains BC-416) read out with eight photomultipliers, two on each end of four 89-cm-long 1.27 cm x 1.27 cm square-cross-section wavelength shifter bars (St. Gobains BC-482) mounted with a small air gap by the edges of the scintillator. A Hamamatsu R1924 photomultiplier is mounted to each end of each bar.

Each of the Cherenkov detectors consists of a 116 cm x 116 cm square box, 21 cm high, equipped with six Burle S83006F photomultipliers looking into the box along each of the four sides. In the first of the Cherenkov detectors, the radiator is composed of a mosaic of four 3-cm-thick square panes of silica aerogel with active area 51 cm x 51 cm and index of refraction 1.04. In the second Cherenkov detector the radiator is a 1.06-cm thick 114 cm x 114 cm square sheet of acrylic (index of refraction 1.5).

Each of the two hodoscope planes consists of two perpendicular layers of square-cross-section fibers. Each fiber is a 1-mm x 1-mm polystyrene scintillator (index of refraction 1.62) surrounded by 0.04-mm acrylic cladding (index of refraction 1.50). The fibers are formatted into tabs of six or seven fibers, so the effective segmentation of the hodoscope is ~6 to 7 mm. The fibers at one end of each layer are grouped into fourteen "segments" with fourteen adjacent tabs forming one segment and going to one Hamamatsu R1924 photomultiplier. The fibers at the other end are grouped with the first tab of each segment going to one photomultiplier, the second tab of each segment going to another photomultiplier, etc. This coding scheme allows us to identify uniquely which of the 196 tabs in a layer is illuminated, using only 28 PMTs per hodoscope (Lawrence et al. 1999).

The TIGER instrument is described in more detail by Link (2003) and Rauch (2008). Short descriptions of the instrument and preliminary reports of its results were given by Sposato et al. (2000), Link et al. (2001, 2003), Geier et al. (2003), Geier et al. (2005), de Nolfo et al. (2005), Rauch et al. (2007), and de Nolfo et al. (2007).

## 3. DATA ANALYSIS

The charge, Z, of each cosmic-ray nucleus that penetrated the instrument was determined from a combination of signals from the two upper scintillator detectors and the two Cherenkov detectors. For each detector the signal was taken as the sum of the signals from all its photomultipliers. That sum was corrected for photomultiplier gain differences, area nonuniformities, and temporal variations, determined by mapping the responses to $\sim 10^6$ cosmic-ray $_{26}$Fe nuclei. The particle trajectory determined from the hodoscope data was used to locate the particle for making the area corrections and to determine the angle ($\theta$) of the trajectory with respect to the normal in order to account for the sec $\theta$ dependence of signal on path-length in the detectors. Particles that suffered a charge-changing nuclear interaction in the instrument were eliminated by requiring agreement to within approximately one charge unit between signals in the upper and lower scintillators.

Figure 2 is a crossplot in which each point represents the two Cherenkov signals of a single cosmic-ray nucleus. The 1.04 index of refraction of the aerogel sets a 2.5 GeV/nucleon threshold for useful signals from the C0 detector. Above this energy, the combination of the two Cherenkov signals gives a well resolved assignment of nuclear charge. Points in Figure 2 at low values of C0 signal, below the diagonal line, represent events due to nuclei with energy below or very close to the C0 threshold; those lower-energy events are analyzed using the signals from the scintillators and the acrylic Cherenkov.

Figure 3 shows a similar cross-plot of signal in the acrylic Cherenkov vs. scintillator signal. The events below the diagonal line have energy above the 2.5 GeV/nucleon aerogel threshold; the charges derived from scintillator and acrylic signals are not well resolved for these high-energy events, but those are the events whose charges are well determined from the two Cherenkov signals. The region to the left of the diagonal line has particles with energy between 350 MeV/nucleon (the acrylic Cherenkov threshold) and 2.5 GeV/nucleon. Here charge is well determined by the combination of scintillator and acrylic signals. Nuclei with energy below 350 MeV/nucleon are identified by having signals to the left of the diagonal edge in the data at low C1 signal, and are not shown.

Figure 4 shows the resulting charge histogram, combining data from the fifty days of the two flights, with charge resolution of $\sigma = 0.23$ charge units. The data display well-defined element peaks at Z = 31, 32, and 34. At higher charges the statistics are inadequate to demonstrate resolution conclusively; however, there is a good suggestion of a peak at Z=38. The smooth curve through the data is a multi-Gaussian maximum-likelihood fit. This fit is used to derive the relative abundances shown in Table 1 for the elements, $_{26}$Fe, $_{28}$Ni, $_{30}$Zn, $_{31}$Ga, $_{32}$Ge, $_{34}$Se, $_{38}$Sr, and the upper limit for $_{33}$As. Since the abundances of $_{27}$Co and $_{29}$Cu are so low relative to their adjacent elements, with Co/Fe and Cu/Ni both of order 1%, we derive the abundances of Co and Cu from a more selected data set, shown in Figure 5. For this data, set the requirement for charge consistency for the Fe-Ni region was increasingly tightened until the Co/Ni ratio remained constant as a function of charge consistency criteria, ensuring minimal contamination of the $_{27}$Co and $_{29}$Cu distributions. It was possible to use this tighter selection to derive Co and Cu abundances, but not for $_{31}$Ga or heavier elements because both Co and Cu are much more abundant than Ga and heavier elements.

The fit to the data of Figure 4 gives a $_{27}$Co/$_{26}$Fe ratio of 0.013 and a $_{29}$Cu/$_{28}$Ni ratio of 0.015, while those two element ratios derived from the higher-resolution data set of Figure 5 are 0.009 and 0.011 respectively. Thus, we infer that with the less selective data set of Figure 4, the spill-over fraction from an abundant even-charge peak to the adjacent odd-charge peak above it is approximately 0.004 of the even-charge peak. The fit to the data of Figure 4 gives a $_{31}$Ga/$_{30}$Zn ratio of 0.129, but we have applied this correction of 0.004 to derive our Ga/Zn ratio at the detector of 0.125.

We have derived abundances at the top of the atmosphere by correcting for the charge-dependent probability of losing particles to interactions in the instrument, the charge-dependent energy loss in the atmosphere, and nuclear interactions in the atmosphere. Both the total and the partial charge-changing cross sections were derived from the accelerator data of Nilsen et al. (1995). Details of this process are described by Rauch (2008). The resulting abundances at the top of the atmosphere are listed in Table 1 and plotted in Figure 6. That figure also shows abundances of $_{30}$Zn, $_{31}$Ga, and $_{32}$Ge measured in space by the HEAO-C2 instrument (Byrnak et al. 1983). Our TIGER results are consistent with those of HEAO-C2 but have lower uncertainties, primarily because of better statistics. That figure also compares these cosmic-ray abundances with the elemental abundances in the solar system (Lodders 2003). (Although Table 1 and Figure 6 give results for $_{35}$Br, $_{36}$Kr, and $_{37}$Rb, the poor statistics and lack of well-resolved peaks give us little confidence in these values and they are not used in the following discussion in Section 4.)

Aside from the well-established depletion of volatile elements relative to refractory elements discussed below, the most striking feature of the data in Figure 6 is that $_{31}$Ga has an abundance about equal to that of $_{32}$Ge, whereas in the solar system the ratio is only ~0.3. As described above, the surprisingly high $_{29}$Ga abundance cannot be explained as spill-over from the more abundant adjacent $_{30}$Zn, since our observed Ga/Zn ratio is an order of magnitude greater than our observed ratios of $_{27}$Co/$_{26}$Fe or $_{29}$Cu/$_{28}$Ni.

4. DISCUSSION

Elemental abundances at the cosmic-ray source were derived from the TIGER abundances at the top of the atmosphere using a leaky-box propagation model (Wiedenbeck et al. 2007), which used partial cross sections based on Silberberg, et al. (1998) and total destruction cross-sections derived from work of Webber et al. (1990). The interstellar propagation results were used as inputs to a spherically symmetric modulation model (Fisk, 1971) with a modulation level $\phi = 850$ MV to obtain modulated values for comparison with the observations made near Earth. The assumed cosmic-ray source abundances were adjusted to yield agreement with the data. The resulting cosmic-ray source abundances are listed in the right-most column of Table 1. (The value of $\phi$ used in these calculations was inferred from the spectra observed by the ACE/CRIS instrument (Wiedenbeck et al., 2005) at the times of these TIGER flights. The resulting source abundances were very insensitive to the value of $\phi$.)

Figure 7 plots the ratio of Galactic cosmic-ray source (GCRS) abundances to solar-system (SS) abundances as a function of the first ionization potential of the elements. For elements with Z > 26 the source abundances are those presented in this paper from TIGER. For elements with Z < 26, the source abundances are those derived from HEAO-C2 data by Engelmann et al. (1990). The solar system data are from Lodders (2003).

The plotted error bars here and in the following figures indicate uncertainty of the GCRS abundances. Some SS abundances are currently under discussion; in particular, the abundances of light elements such as O and Ne might be underestimated (Basu & Anita, 2008). As previously noted, there is a general trend of lower GCRS/SS for elements with higher first ionization potential, but there is considerable scatter in the points, suggesting that first ionization potential is not directly controlling the abundances. (If instead of comparing the GCRS abundances with SS abundances, we were to use the mix of SS and outflow of massive stars described below, the scatter would be even worse. For example the point for $_{31}$Ga, which has a first ionization potential of 6.0 eV, would move from an ordinate above 1 down to 0.4.)

Figure 8 plots the same GCRS/SS ratios, as a function of elemental atomic mass, indicating separately refractory elements (with equilibrium condensation temperature (Lodders 2003) greater than ~1200 K) and volatile elements (with lower condensation temperature). Here we observe two effects noted by Meyer et al. (1997) and Ellison et al. (1997). The GCRS/SS ratio is generally higher for refractory elements than for volatile elements, and the value of this ratio for volatile elements shows a power-law trend as a function of atomic mass. We note, however, that there remains significant scatter, suggesting that this is not the full picture.

As noted above, the most striking feature of the new data presented here is the high abundance of $_{31}$Ga, evident in both Figures 6 and 8. We note that in an analysis by Woosley & Heger (2007) of the initial-mass-function-integrated yield of outflow from massive stars, the element that has the greatest overabundance compared with solar-system composition is Ga. Since there is evidence that the cosmic-ray source composition is a mixture of material with solar-system composition and material typical of outflow from massive stars (Higdon & Lingenfelter 2003, Binns et al. 2005), Figure 9 is a modification of the previous figure where now the ordinate is the ratio of GCRS to a mixture by mass of 80% solar-system composition (Lodders 2003) and 20% initial-mass-function-integrated yield of outflow from massive stars including both their Wolf-Rayet and supernova phases (Woosley & Heger 2007). This figure shows a significant improvement in the organization of the data compared with Figure 8. Here the refractory elements have a mass (A) dependence of approximately $A^{2/3}$, and volatile elements have a mass dependence of approximately $A^1$. Similarly good organization of the data is found with any mixtures in which the fraction of massive-star outflow has a value between about 15% and 25%. Thus the precise value of the mixture is not tightly constrained by this analysis, although it is clear that some fraction of massive-star outflow is required. This range of 15% to 25% is similar to the result of (18 ± 5)% found by Higdon & Lingenfelter (2003) and a similar range of 12% to 25% derived by Lingenfelter and Higdon (2007).

We have used here the same division of elements into volatile and refractory groups as was used in Figure 8. The condensation temperatures for a gas with an overall composition of the 80%/20% mixture are similar to those calculated for a solar composition gas, which is not too surprising because its C/O ratio remains below unity and the minerals condensing from it are essentially the same as those in a gas of solar composition.

Note that the oxygen datum in Figure 9 was ignored in the fit to the volatile elements. Although one might naively think of oxygen as volatile, in fact a significant fraction of the oxygen can be found in dust grains, as has been pointed out by both Meyer et al. (1997) and Lingenfelter et al. (1998). The fraction of oxygen in grains is estimated from the abundances of the elements like Mg, Al, Si, and Ca that form high-temperature oxide condensates. With a solar-system composition 23% of the total oxygen would thus condense (Lodders, 2003), but with the composition of the 80%/20% mixture only 12% of the total oxygen would be sequestered in grains. In fact, the difference between the ordinate of the oxygen point in Figure 9 and the volatile fit line is very nearly 12% of the difference between the refractory and volatile lines at mass 16. Thus, the oxygen datum in this figure is consistent with the model of refractory/volatile differentiation of GCR source abundances.

We take Figure 9 as strong evidence in support of the model of cosmic-ray origin in OB associations and the model of cosmic-ray acceleration in which elements found in interstellar grains are more effectively accelerated than are elements found in the interstellar gas.

ACKNOWLEDGEMENTS

We gratefully acknowledge the excellent work of the staff of the Columbia Scientific Balloon Facility, the NASA Balloon Program Office, and the NSF Office of Polar Programs who together made possible two successful long-duration balloon flights over Antarctica. This research was supported by NASA at Washington University (grant NNG05WC04G), Goddard Space Flight Center, the California Institute of Technology (grant NNG05WC21G), and Caltech's Jet Propulsion Laboratory. K.L.'s participation at Washington University was supported by the NSF (grant AST0807356).

FIGURE CAPTIONS

Figure 1.  TIGER instrument cross section

Figure 2.  Aerogel Cherenkov signal vs. Acrylic Cherenkov signal.  For Z>27.5, every event in the 2003 data set is plotted.  For 14.5<Z<27.5, only 1/15 are plotted.  For Z<14.5, only 1/30 are plotted.  Events below the line have energy below or very close to the C0 threshold (2.5 GeV/nucleon).

Figure 3.  Signal in acrylic Cherenkov detector vs. scintillator signal.  For Z>27.5, every event in the 2003 data set is plotted.  For 14.5<Z<27.5, only 1/15 are plotted.  For Z<14.5, only 1/30 are plotted.  Events to the right of the line have energy above the C0 threshold (2.5 GeV/nucleon).

Figure 4.  Combined data from both flights with maximum-likelihood multi-Gaussian fit.

Figure 5.  High-resolution data set from the 2001 flight.  The maximum-likelihood fit to these data was used to determine the abundances of $_{27}$Co and $_{29}$Cu.

Figure 6. Cosmic-ray abundances compared with solar system abundances. Solid squares are TIGER data corrected to the top of the atmosphere. Open circles are results from HEAO-C2 (Byrnak et al. 1983). Histogram is solar-system abundances (Lodders 2003).

Figure 7. The ratio of galactic cosmic-ray source (GCRS) abundances to solar-system (SS) abundances vs. first ionization potential of the elements.

Figure 8. The ratio of galactic cosmic-ray source (GCRS) abundances to solar-system (SS) abundances vs. atomic mass (A). Refractory elements are shown as blue circles; volatile elements, as red squares.

Figure 9 This figure is the same as figure 8, except the reference abundances to which galactic cosmic-ray source (GCRS) abundances are compared are a mixture of 80% solar system (SS) and 20% from massive star outflow (MSO). Note that oxygen is not treated as either volatile or refractory.

Table 1 Element abundances relative to $_{26}$Fe

| Z | | Observed at TIGER | Top of Atmosphere | Source |
|---|---|---|---|---|
| 27 | Co | $(1.01 \pm 0.03) \times 10^{-2}$ | $(0.81 \pm 0.04) \times 10^{-2}$ | $(0.44 \pm 0.04) \times 10^{-2}$ |
| 28 | Ni | $(4.93 \pm 0.03) \times 10^{-2}$ | $(5.16 \pm 0.03) \times 10^{-2}$ | $(5.39 \pm 0.03) \times 10^{-2}$ |
| 29 | Cu | $(5.4 \pm 0.7) \times 10^{-4}$ | $(5.6 \pm 0.8) \times 10^{-4}$ | $(5.5 \pm 0.9) \times 10^{-4}$ |
| 30 | Zn | $(5.6 \pm 0.3) \times 10^{-4}$ | $(6.2 \pm 0.4) \times 10^{-4}$ | $(6.5 \pm 0.4) \times 10^{-4}$ |
| 31 | Ga | $(7.0 \pm 1.1) \times 10^{-5}$ | $(7.3 \pm 1.3) \times 10^{-5}$ | $(6.9 \pm 1.5) \times 10^{-5}$ |
| 32 | Ge | $(6.0 \pm 1.1) \times 10^{-5}$ | $(6.7 \pm 1.3) \times 10^{-5}$ | $(5.5 \pm 1.5) \times 10^{-5}$ |
| 33 | As | $(9.1 \pm 4.4) \times 10^{-6}$ | $< 1.0 \times 10^{-5}$ | $< 0.7 \times 10^{-5}$ |
| 34 | Se | $(5.5 \pm 1.1) \times 10^{-5}$ | $(6.7 \pm 1.4) \times 10^{-5}$ | $(7.1 \pm 1.7) \times 10^{-5}$ |
| 35 | Br | $(1.0 \pm 0.5) \times 10^{-5}$ | $(0.9 \pm 0.7) \times 10^{-5}$ | $< 1.3 \times 10^{-5}$ |
| 36 | Kr | $(1.7 \pm 0.6) \times 10^{-5}$ | $(1.9 \pm 0.8) \times 10^{-5}$ | $(1.3 \pm 1.0) \times 10^{-5}$ |
| 37 | Rb | $(1.1 \pm 0.5) \times 10^{-5}$ | $(1.2 \pm 0.7) \times 10^{-5}$ | $(1.3 \pm 0.8) \times 10^{-5}$ |
| 38 | Sr | $(2.2 \pm 0.7) \times 10^{-5}$ | $(3.0 \pm 1.0) \times 10^{-5}$ | $(3.7 \pm 1.2) \times 10^{-5}$ |

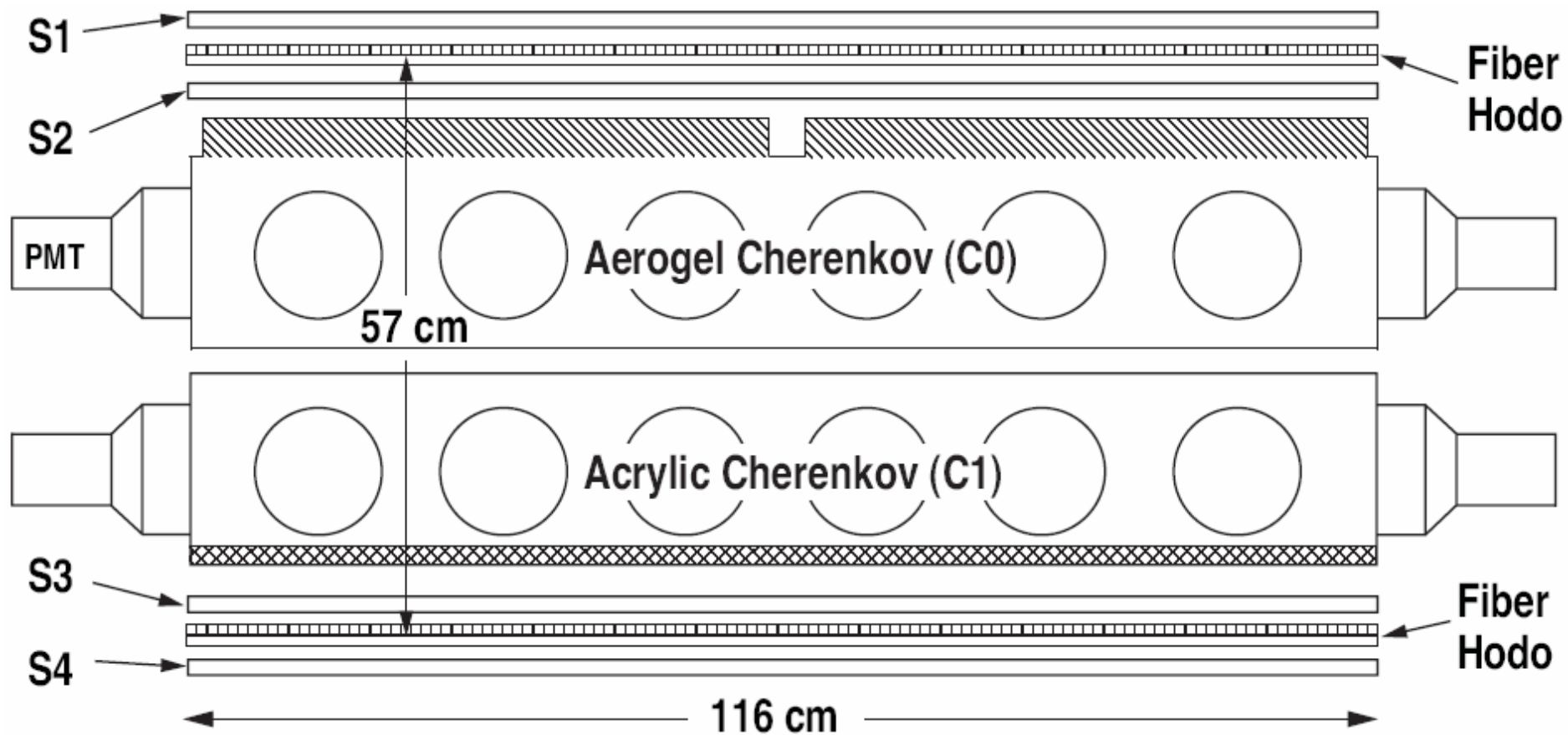

Figure 1

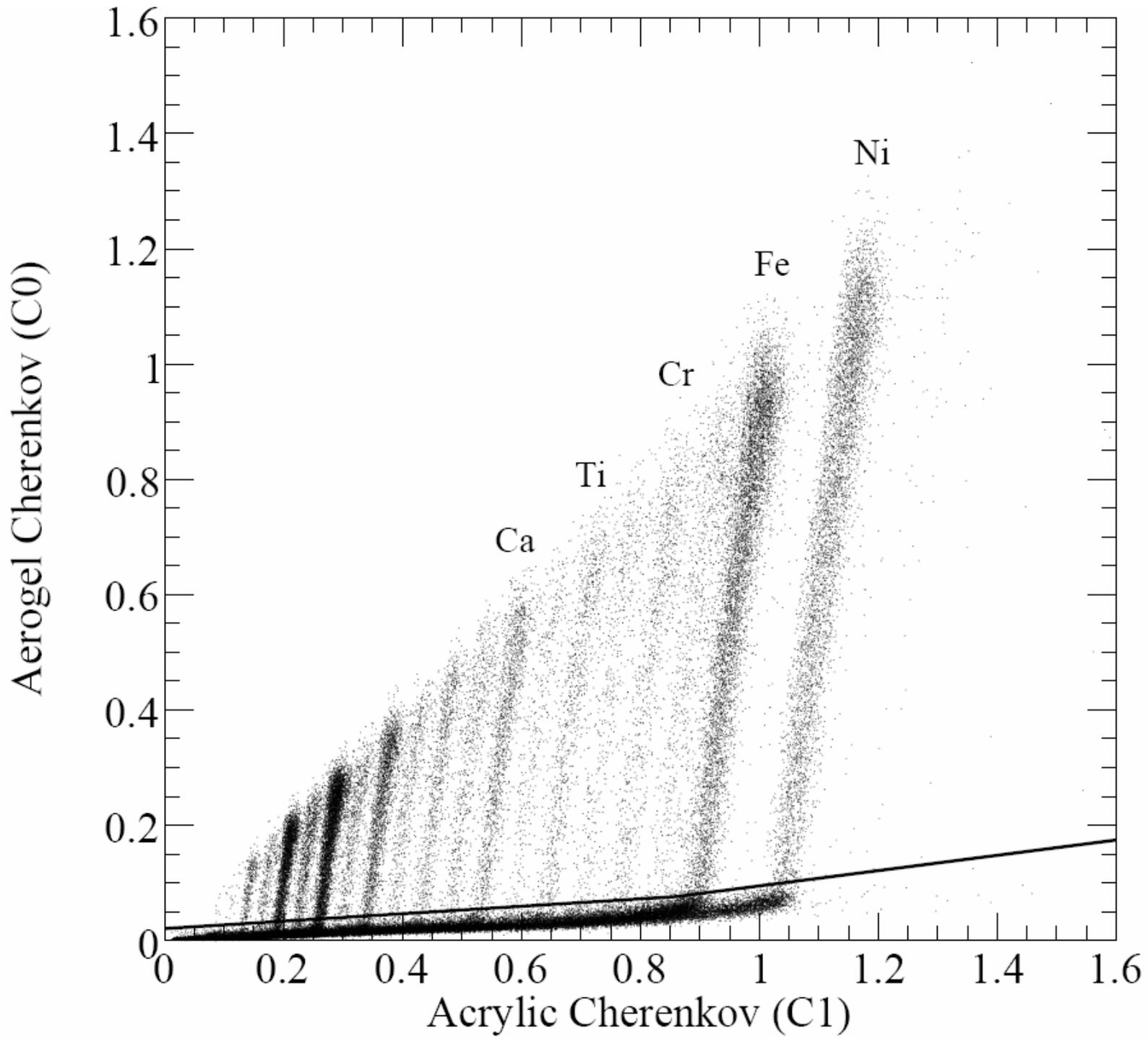

Figure 2

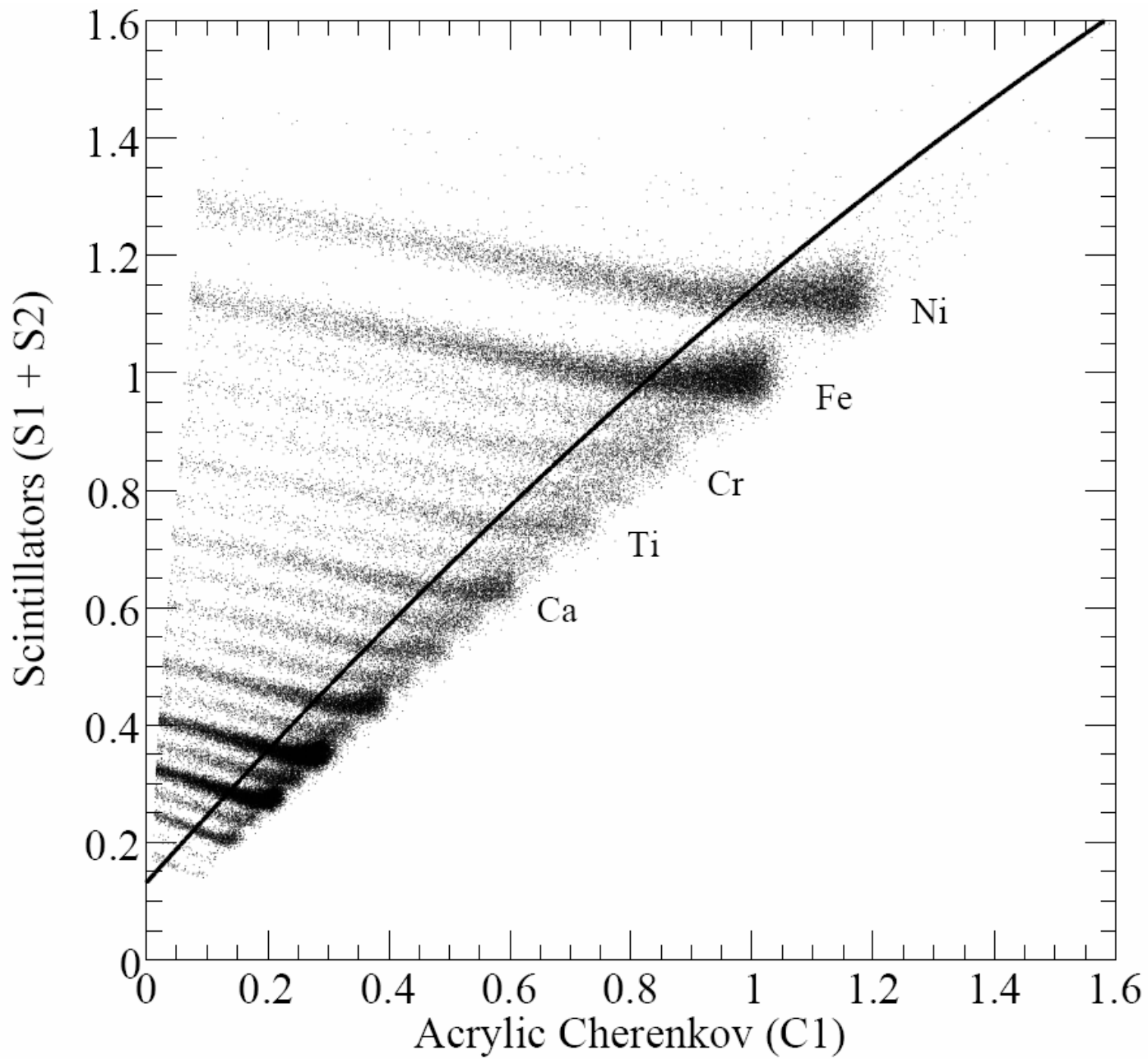

Figure 3

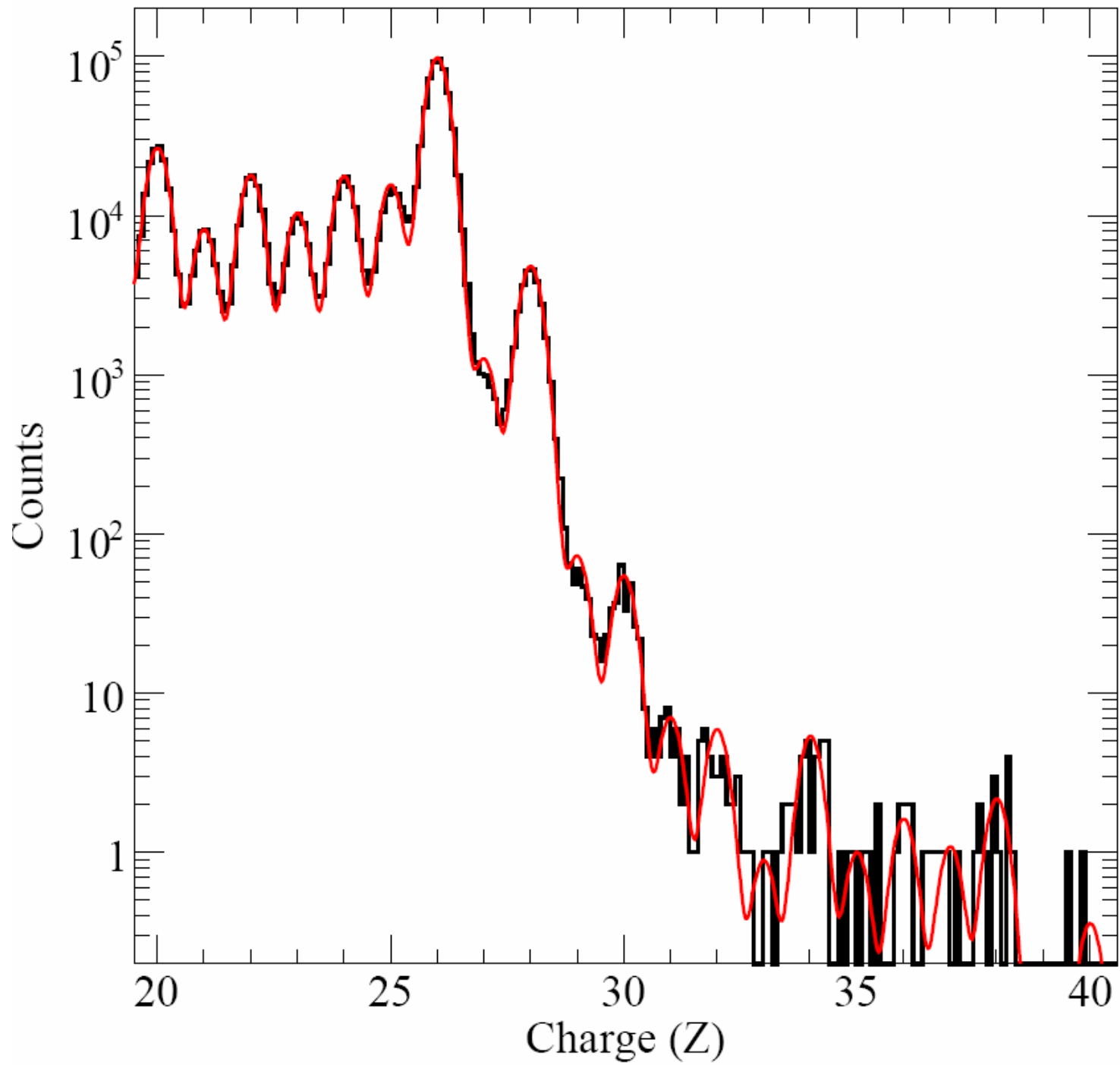

Figure 4

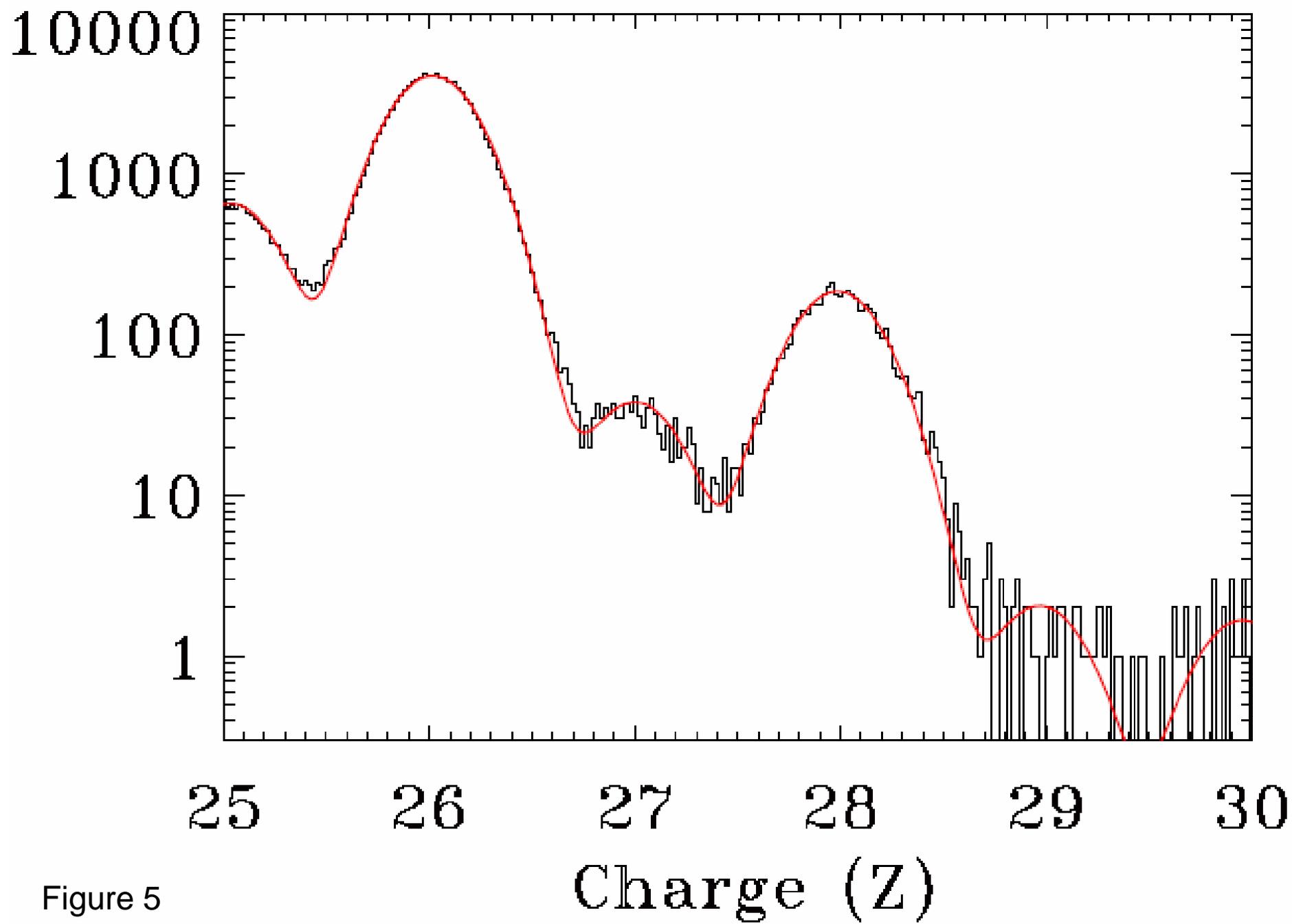

Figure 5

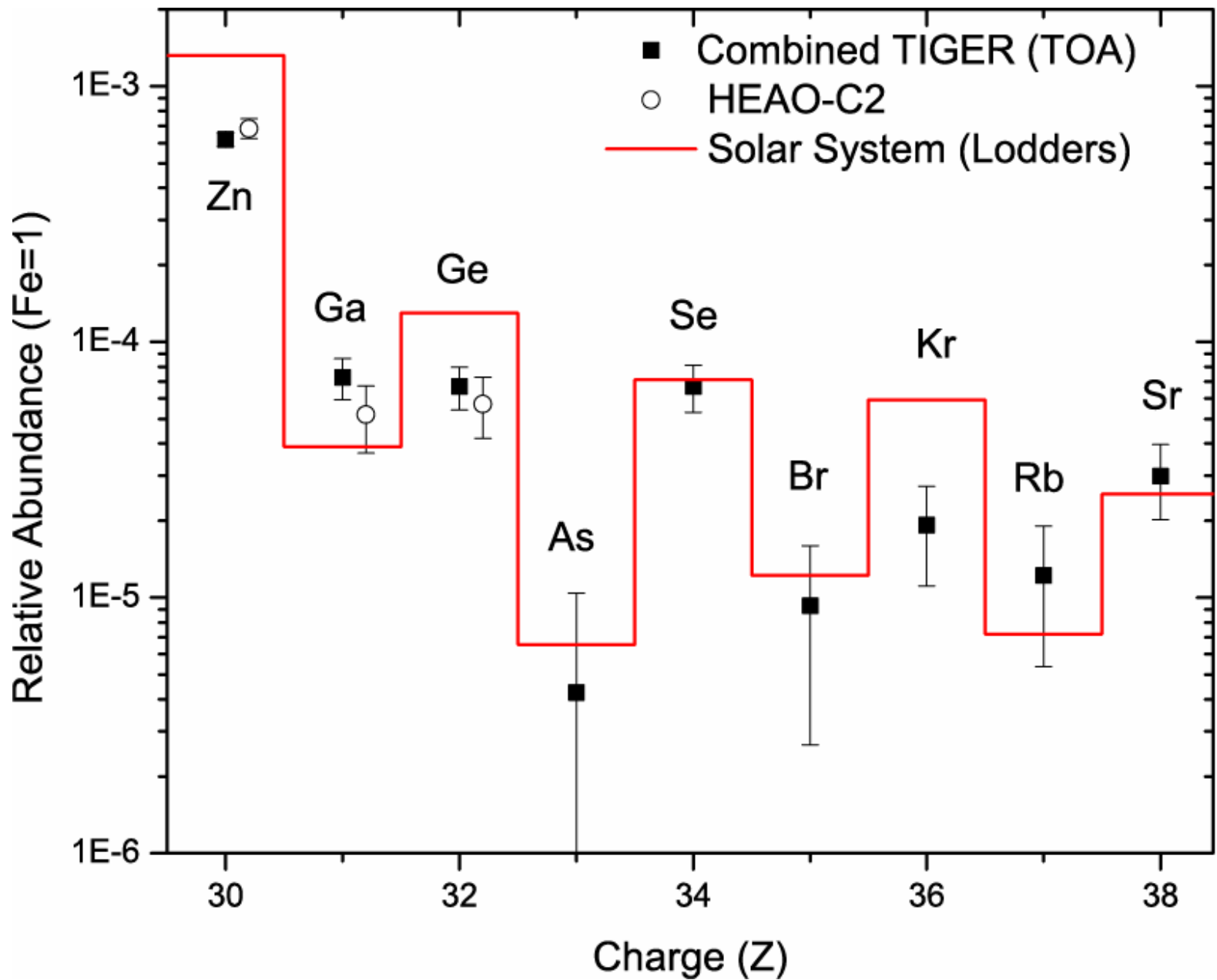

Figure 6

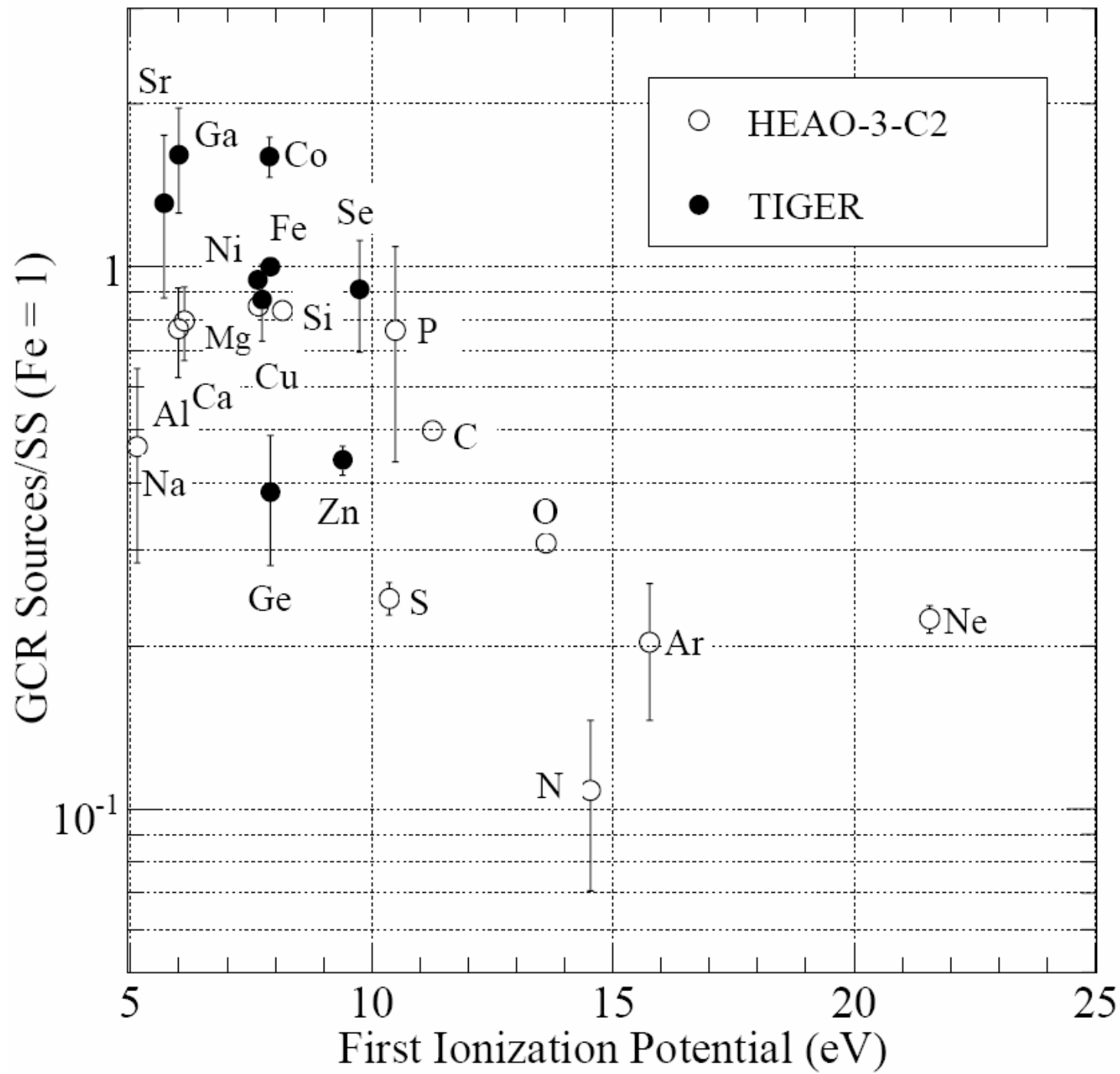

Figure 7

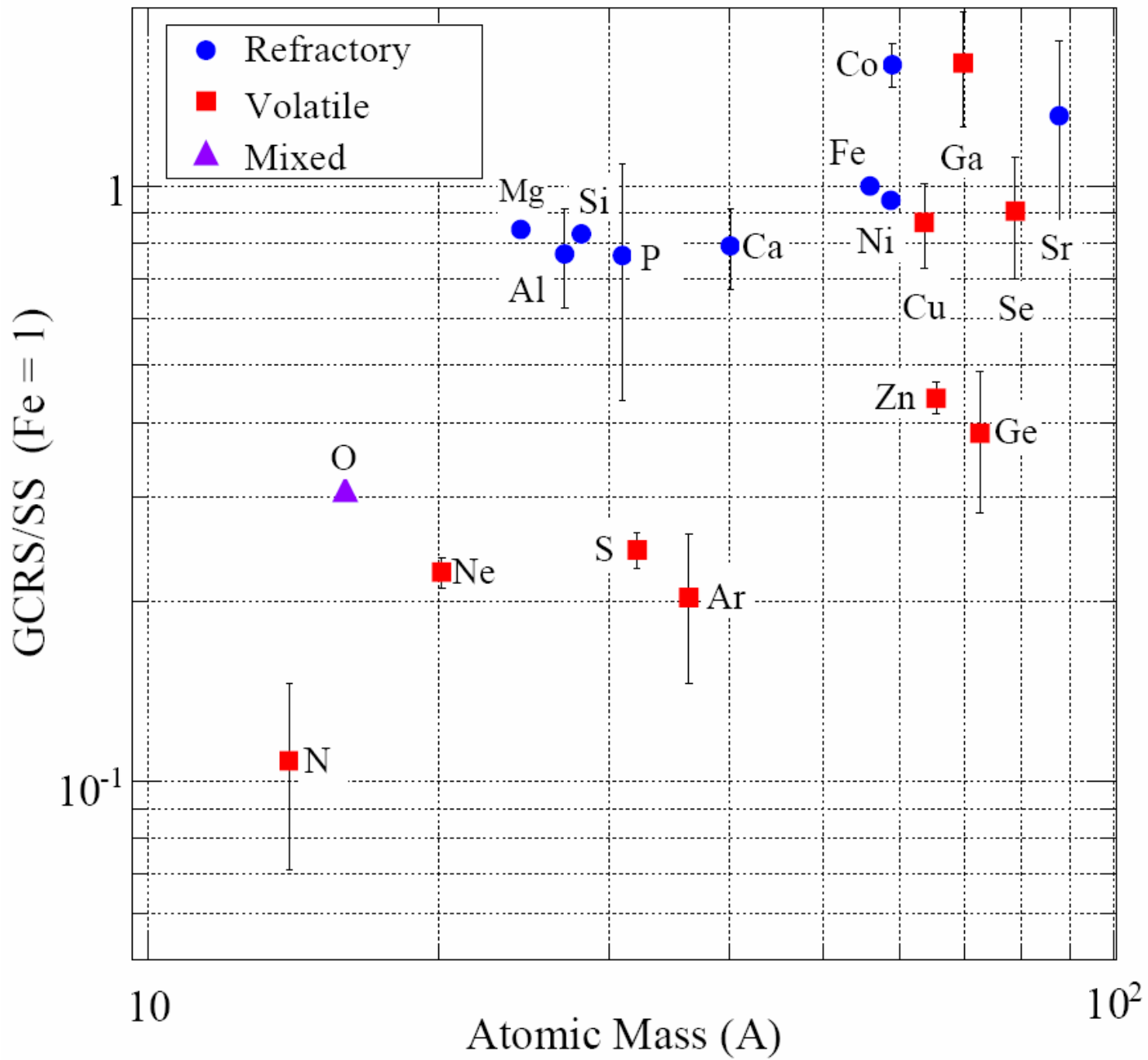

Figure 8

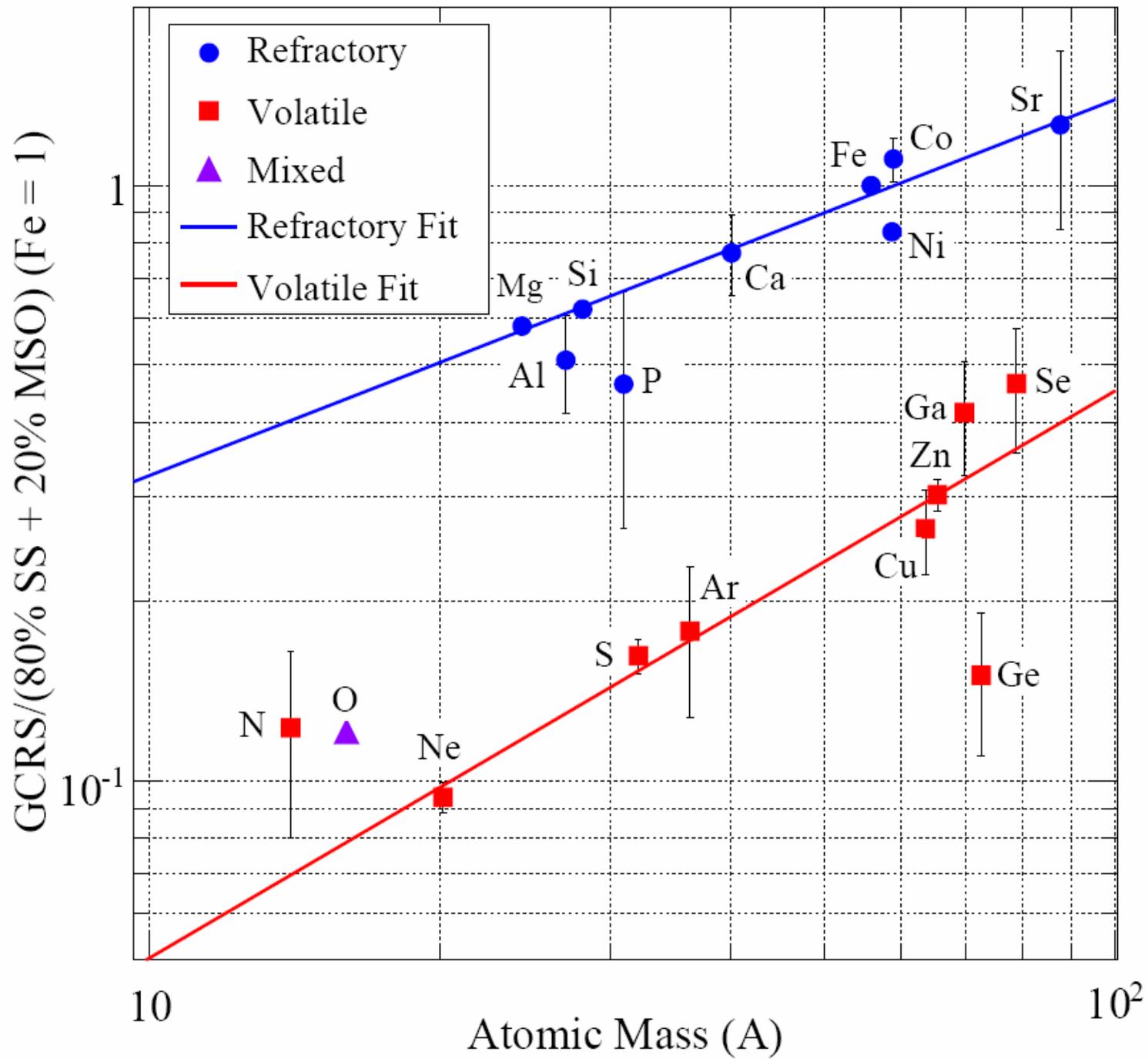

Figure 9